\documentclass[conference]{IEEEtran}
\IEEEoverridecommandlockouts
\usepackage{cite}
\usepackage{amsmath,amssymb,amsfonts}
\usepackage{algorithmic}
\usepackage{graphicx}
\usepackage{textcomp}
\usepackage{xcolor}
\def\BibTeX{{\rm B\kern-.05em{\sc i\kern-.025em b}\kern-.08em
    T\kern-.1667em\lower.7ex\hbox{E}\kern-.125emX}}

\usepackage{hyperref}
\hypersetup{
    colorlinks=true,
    linkcolor=blue,     
    citecolor=blue,     
    filecolor=blue,     
    urlcolor=blue       
}
\usepackage{booktabs} 

\begin{document}

\title{Simultaneous Music Separation and Generation Using Multi-Track Latent Diffusion Models\\
\thanks{This work was supported by IRCAM and Project REACH (ERC Grant 883313) under the EU’s Horizon 2020 programme.}
}

\author{
\IEEEauthorblockN{Tornike Karchkhadze$^{\dag}$}
\IEEEauthorblockA{\textit{University of California San Diego} \\
San Diego, USA  \\
tkarchkhadze@ucsd.edu}
\and
\IEEEauthorblockN{Mohammad Rasool Izadi$^{\dag}$}
\IEEEauthorblockA{\textit{Bose Corp.} \\
Framingham, USA \\
russell\_izadi@bose.com
}
\and
\IEEEauthorblockN{Shlomo Dubnov}
\IEEEauthorblockA{\textit{University of California San Diego} \\
San Diego, USA \\
sdubnov@ucsd.edu
}


}

\maketitle

\def\thefootnote{\dag}
\footnotetext{Authors with equal contribution.}

\begin{abstract}
Diffusion models have recently shown strong potential in both music generation and music source separation tasks. Although in early stages, a trend is emerging towards integrating these tasks into a single framework, as both involve generating musically aligned parts and can be seen as facets of the same generative process. In this work, we introduce a latent diffusion-based multi-track generation model capable of both source separation and multi-track music synthesis by learning the joint probability distribution of tracks sharing a musical context. Our model also enables arrangement generation by creating any subset of tracks given the others. We trained our model on the Slakh2100 dataset, compared it with an existing simultaneous generation and separation model, and observed significant improvements across objective metrics for source separation, music, and arrangement generation tasks. Sound examples are available at \href{https://msg-ld.github.io/}{https://msg-ld.github.io/}.
\end{abstract}


\begin{figure*}[t]
  \centering
  \includegraphics[width=.95\linewidth]{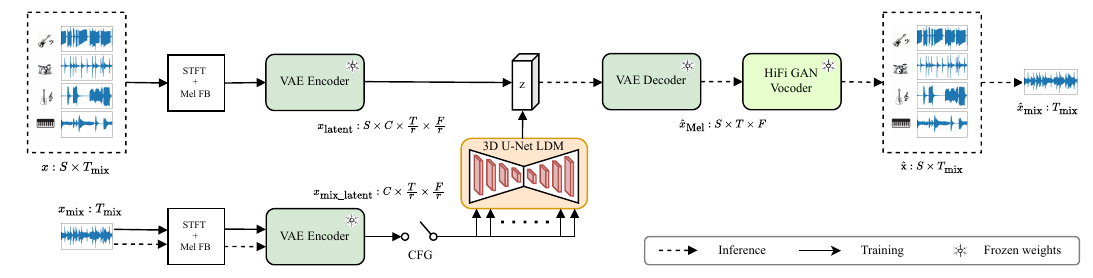}
  \caption{MSG-LD system overview: During training, audio tracks are converted into Mel-spectrograms and compressed into a 3D latent space by a VAE encoder, where LDM operates. The audio mixture is similarly processed and used as a condition by adding it to each U-Net layer. During inference, the model's conditioning is controlled by CFG weight, switching between source separation and music generation modes. The generated latent vectors are up-sampled to Mel-spectrograms by the VAE decoder and converted into audio via HiFi-GAN vocoder.}
  \label{fig1}
\end{figure*}

\section{Introduction}\label{sec:introduction}

Audio source separation refers to the process of isolating individual sound elements from a mixture of sounds. This method is critical in numerous areas, particularly in music production. Recent advances with deep learning models have significantly improved separation quality, utilizing two different approached: discriminative
\cite{choi2021lasaft, défossez2022hybrid, Lluis2019, guso2022loss,luo2019conv, defossez2019music, takashi2018mmdenselstm} and generative \cite{subakan2018generative, kong2019single, narayanaswamy2020unsupervised, zhu2022music, jayaram2021parallel, jayaram2020source, manilow2022improving, postolache2023adversarial, kavalerov2019universal, wisdom2020unsupervised}. 
Discriminative models focus on directly mapping the input mixture to its separated sources, whereas generative models aim to learn the distribution of individual sources and their combination into a mixture. Diffusion models \cite{ho2020denoising} have lately become a leading generative method in audio source separation tasks \cite{scheibler2023diffusion, lutati2023separate, Huang2023DAVIS, Plaja-RoglansMS23, mariani2024multisource}. 

On the other hand, music generation in the domain of raw audio has seen significant advancements with the development of deep learning techniques \cite{vandenoord2016wavenet, Mehri2016, donahue2018adversarial, dhariwal2020jukebox, agostinelli2023musiclm, MusicGen2023}. These systems create audio content either unconditionally or conditioned on various modalities, often focusing on generating music based on text descriptions of musical genres, mood, and other attributes. Having demonstrated their ability to learn complex data distributions, such as raw audio, diffusion models have had a profound impact on music generation in the audio domain~\cite{chen2023musicldm, melechovsky2024mustango,  schneider2023mousai, kong2021diffwave}. An alternative paradigm for music generation, which involves using existing musical tracks or melodic hints to generate the remaining music in response to the musical context, has been explored in music-to-music and musical arrangement generation models \cite{yao2023jen1, stemgen2024, pasini2024bass, donahue2023singsong, karchkhadze2024multitrack, nistal2024diffariff}.



In the deep learning literature, music separation and generation have traditionally been treated independently \cite{Moysis23}. Typically, music generation models—whether conditioned or unconditioned—aim to learn the distribution of the entire mixture of sounds, which makes source separation unfeasible. Conversely, source separation models isolate individual sources but lose critical information about the mixture, thereby hindering the possibility of full music generation. Unlike other audio fields, in music, sound is often a composite of tightly interdependent tracks. Learning the joint distribution of these tracks can be useful for both separation and generation tasks, which can be viewed as related. These two tasks can be seen as point on a spectrum—from unconditional music generation, to generating tracks conditioned on the mixture, fully decomposing it into individual components. Recent work, such as the Multi-Source Diffusion Model (MSDM)~\cite{mariani2024multisource}, has demonstrated the feasibility of addressing both music generation and source separation within a unified framework.



We introduce the Music Separation and Generation with Latent Diffusion (MSG-LD) model, which learns the joint probability of latent representations of interrelated musical tracks composing mixtures. Our model performs three tasks: In the case of \textit{Source Separation}, it decomposes a conditioned mixture into individual tracks. In unconditional mode, it performs \textit{Total Generation}, creating new compositions across multiple tracks. Using inpainting, it also handles \textit{Partial Generation} or arrangement generation, producing missing tracks based on others, such as, for example, adding a guitar to an existing bass and drums tracks.

As part of this approach, we utilized the MusicLDM~\cite{chen2023musicldm}, an adaptation of AudioLDM~\cite{pmlr-v202-liu23f} for music, and extended it into a conditional multi-track audio diffusion model.
To enable flexible control and seamlessly alternate between separation and generation tasks, we employed the Classifier-Free Guidance (CFG)~\cite{ho2022classifierfree} paradigm to adjust conditioning strength.
Our experiments demonstrate that the model produces realistic music across various scenarios: source separation, total track-by-track music generation, and arrangement generation with any combination of tracks. 
We compared our model to MSDM, the only other model known to handle these tasks simultaneously, and used it as our baseline.
Compared to MSDM, our model, trained on the same datataset Slakh2100~\cite{Slakh}, achieves significant improvements in all tasks based on objective evaluations.
As part of our commitment to reproducibility and open science, the code and checkpoints of this study are publicly available \footnote{\href{https://github.com/karchkha/MSG-LD}{https://github.com/karchkha/MSG-LD}}.




\section{Method}
\label{Method}


Depicted in Fig.~\ref{fig1}, our proposed system, MSG-LD builds on the foundation of MusicLDM~\cite{chen2023musicldm} and utilizes the Latent Diffusion Model (LDM)~\cite{rombach2022high, ho2020denoising} as its framework. We replaced the original text conditioning with mixture audio conditioning and extended LDM generator to handle multiple tracks simultaneously.


Let \( x_{\text{mix}} \) represent a time-domain audio mixture consisting of \( S \) tracks \( x_s \), where \( s \in \{1, \dots, S\} \) and the duration is \( T_{\text{mix}} \). The mixture is defined as \( x_{\text{mix}} = \sum_{s=1}^S x_s \). The stack of individual waveforms is denoted as \( x \in S \times T_{\text{mix}} \). As shown in Fig.~\ref{fig1}, \( x \) is processed via short-time Fourier transform (STFT) and Mel operations, resulting in a Mel-spectrogram \( x_{\text{Mel}} \in S \times T \times F \), where \( T \) and \( F \) represent time and frequency. The encoder of a pretrained Variational Autoencoder (VAE)~\cite{kingma2022autoencoding} then compresses \( x_{\text{Mel}} \) into a latent representation \( x_{\text{latent}} \in S \times C \times \frac{T}{r} \times \frac{F}{r} \), where \( r \) is the VAE's compression ratio, and \( C \) the number of latent channels. We denote this space as \( z = x_{\text{latent}} \). In this space, the model learns the distribution \( q(z) \) under the LDM framework. Finally, the VAE decoder reconstructs \( \hat{z} \) back to the Mel-spectrogram \( \hat{x}_{\text{Mel}} \), which is converted to time-domain audio \( \hat{x} \) using pretrained HiFi-GAN vocoder~\cite{kong2020hifi}.

\subsection{Multi-Track LDM}

We use denoising diffusion probabilistic models (DDPMs)~\cite{ho2020denoising, Sohl-DicksteinW15} as our generator. DDPMs are a class of generative models that mimic the thermodynamic process of diffusion to generate data. DDPMs add and remove controlled amounts of noise from data over a series of time steps in a forward and reverse process. The forward pass gradually introduces Gaussian noise, \(\epsilon \sim \mathcal{N}(0, I)\), to the clean latent variable \(z_0\) as \(z_n = z_0 + \sigma_n \epsilon\), where \(\sigma_n\) controls the noise scheduling at each step \(n \in \{1, \ldots, N\}\), with \(N\) being the total number of steps. This process ultimately results in isotropic Gaussian noise \(z_N \sim \mathcal{N}(0, I)\). For the reverse process, the model is trained to estimate and remove the added noise at each step, ultimately generating the latent variable \(z_0 \sim q(z_0)\) from Gaussian noise \(z_N\), either conditionally or unconditionally. This process can be represented as a Markovian process \(p_{\theta}(z_0) = \int p_{\theta}(z_{0:N}) dz_{1:N}\), where \(\theta\) corresponds to the model parameters. 

The denoising model is trained by minimizing the mean square error (MSE) between the predicted noise \(\epsilon_{\theta}\) and the actual Gaussian noise \(\epsilon\) at every step, following the classic DDPM~\cite{ho2020denoising} loss function at each step, as follows:

\begin{equation}
L(\theta) = \mathbb{E}_{z_0, \epsilon, n} \|\epsilon - \epsilon_{\theta}(z_n,n, [c]) \|^2
\label{eq:loss}
\end{equation}
where \([c]\) denotes the optional use of conditioning.

In our DDPM model, we utilize a large U-Net~\cite{RonnebergerFB15} architecture as the backbone for the diffusion. To accommodate \(z_n\) with an additional dimension \(S \times C \times \frac{T}{r} \times \frac{F}{r}\), rather than the single-channel audio latent representation \(C \times \frac{T}{r} \times \frac{F}{r}\) used in the original Audio/MusicLDM, we extend the U-Net architecture by incorporating 3D convolutional operations in place of 2D convolutions. This effectively extends the dimensionality of the U-Net to 3D, treating the channel dimension of \(C\) as an additional spatial dimension. Consequently, the track dimension \(S\) now serves as the new channel dimension.

\subsection{Separation and Generation Trade-off}

To enable controllable generation, in DDPM model framework one can introduce a condition \( c \) to the diffusion process, resulting in \( p_{\theta}(z_0|c) \). To enable a control over adherence to conditioning information, DDPM models often utilize classifier-free guidance (CFG)~\cite{ho2022classifierfree}. This is accomplished by randomly omitting the conditioning information during training, allowing the simultaneous training of both conditional and unconditional versions of the model. During inference, the strength of the conditioning is modulated \(\hat{\epsilon} = w\epsilon_{c} + (1 - w)\epsilon_{u}\), where \(w\) is the CFG guidance scale weight balancing the model's conditional \(\epsilon_{c}\) and unconditional \(\epsilon_{u}\) predictions. 


We utilize CFG paradigm to balance the trade-off between source separation and generation. After simultaneously training the unconditional and conditional models, we vary the guidance scale weight \( w \) during inference. By adjusting \( w \) between 0 and 1, we effectively switch between source separation and total music generation modes of our model. 
Notably, our approach differs from posterior approximation methods like Dirac and Gaussian used in MSDM, which we found incompatible with LDM due to the non-linear relationship between mixtures and sources in the latent space.

\subsection{Music Separation}

To achieve successful music source separation, we introduced a condition over the generation process using the latent representation of the mixture \(c = x_{\textit{mix\_latent}}\), resulting in \(p_{\theta}(z_0|x_{\textit{mix\_latent}})\). The mixture latent \(x_{\textit{mix\_latent}}\) is obtained similarly to \(x_{\text{latent}}\) through STFT, Mel operations, and a VAE encoder, as shown in Fig.~\ref{fig1}. 
To ensure strong adherence to the conditioning and avoid "forgetting" the mixture, we replaced FiLM \cite{perez2018film} used in MusicLMD with direct mixture conditioning at every layer of the U-Net.
This is accomplished by processing the latent representation of mixture with down-sampling using average pooling, matching the corresponding sizes of the U-Net layers. Additionally, the channel numbers are also matched by repeating the re-sampled mixture as needed. During inference, we apply a CFG weight \(w \ge 1\) to ensure strong adherence to the conditioning, enabling effective source separation.

\subsection{Music Generation}
After training on multi-track data, the model can generate multiple tracks simultaneously. By learning a joint distribution, it maintains coherence between tracks. We determine two scenarios: total generation and partial generation.

\textbf{Total Generation} creates tracks unconditionally, with a CFG weight \(w = 0\). These tracks can then be used individually or combined to form a musical mixture by summing them. Mathematically, the formation of final music in total generation scenario can be expressed as the summation of the generated \(\hat{x}\) matrix across the first dimension: \( \hat{x}_{\text{mix}} = \sum_{s=1}^S \hat{x}_s \), where each \(\hat{x}_s\) represents the \(s\)-th row of \(\hat{x}\).

\textbf{Partial Generation} involves filling in missing segments of a partially observed multi-track musical piece, akin to arrangement composition in music.  In machine learning literature, particularly in the image domain, this task is commonly known as imputation or inpainting~\cite{Sohl-DicksteinW15, lugmayr2022repaint}. Using the LDM model, the process of arrangement generation for a given latent representation of the tracks, \(z_I = \{ z_s | s \in I \}\), involves finding the latent representation for the missing tracks, \(z_{\bar{I}} = \{ z_s | s \in \bar{I}\}\), with \(\arg\max_{z_{\bar{I}}} p_{\theta}( z_{\bar{I}} | z_I)\), where \(\bar{I} = \{1, \dots, S\} - I\). The inpainting method we employ operates only during inference and does not require any special training. Instead, we leverage an unconditional diffusion model as a generative prior, ensuring harmonization between missing and given parts of the data. Essentially, partial generation becomes a generation problem where, at every step \(n\), the parts of the latent space corresponding to the given tracks are masked and replaced with their noise-added versions, obtained by adding \(n-1\) noise steps to \(z_0\) through the forward process. This approach compels the model to generate under constraints, ensuring that the generated arrangement tracks align well with the provided ones.

\section{Experimental Setup} \label{section_experimental_setup}

\subsection{Dataset}

For our experiments, we used the Slakh2100 dataset~\cite{Slakh}, which contains audio examples synthesized from MIDI files using high-quality virtual instruments. 
To ensure direct comparability with our baseline, MSDM, we employed the same sub-selection of the Slakh2100 dataset, with \(S=4\) of the most prevalent instrument classes: Bass, Drums, Guitar, and Piano.

To align with the specifications of our model, we down-sampled the audios to 16 kHz. We took audio segments of 10.24 seconds, which we extracted from the Slakh tracks with a small random shift for training samples and without adding shifts for validation and testing. We converted the audio clips into Mel-spectrograms with a window length of 1024 and a hop size of 160 samples, resulting in spectrograms of \( F \times T = 64 \times 1024 \). The mixtures were obtained by simply adding individual tracks without using normalization.

\begin{table*}[t]
    \centering
    \caption{FAD\(\downarrow\) Scores for instrument stems (B: Bass, D: Drums, G: Guitar, P: Piano) and their combinations in arrangement generation tasks. The performance of our model is compared to the MSDM baseline. }
    \resizebox{0.85\textwidth}{!}{
        \begin{tabular}{l|cccccccccccccc}
        \toprule
        Model & B & D & G & P & BD & BG & BP & DG & DP & GP & BDG & BDP & BGP & DGP \\
        \midrule
        MSDM \cite{mariani2024multisource} & 0.45 & 1.09 & \textbf{0.11} & 0.76 & 2.09 & 1.00 & 2.32 & 1.45 & 1.82 & 1.65 & 2.93 & 3.30 & 4.90 & 3.10 \\

        MSG-LD & \textbf{0.22} & \textbf{0.78} & 0.29 & \textbf{0.51} & \textbf{0.83} & \textbf{0.59} & \textbf{0.83} & \textbf{0.99} & \textbf{1.11} & \textbf{0.96} & \textbf{1.09} & \textbf{1.15} & \textbf{1.23} & \textbf{1.46} \\

        \bottomrule
        \end{tabular}
    }
    \label{table:arrange_generation}
    \vspace{-5pt}
\end{table*}

\subsection{Model, Training, and Evaluation Setup}

The hyperparameter setting for our model largerly follows MusicLDM~\cite{chen2023musicldm}, with extension of the LDM model by modifying the U-Net architecture to handle a 3D latent space with an additional track numbers dimension \(S=4\). We used a VAE compression ratio \(r = 4\), resulting in a latent space shape of \(S \times C \times \frac{T}{r} \times \frac{F}{r} = 4 \times 8 \times 256 \times 16\), representing stems, channels, time, and frequency, respectively. We used a U-Net channel numbers of 128, 256, 384, and 640 for the encoder and the reversed for the decoder block. Unlike the original Music/AudioLDM, we opted for a simple attention layer instead of a spatial transformer for the U-Net's attention mechanism, as the latter, designed for 2D image-like data, was incompatible with our model's 3D latent space. We used pretrained components of MusicLDM—VAE and the HiFi-GAN vocoder—from publicly available checkpoints. We trained our model with unconditional dropout rate of \(0.1\), Adam optimizer and learning rate of \(3 \times 10^{-5}\) for 100 epochs. The total number of steps for DDPM was set at \(N = 1000\) during training. We used Denoising Diffusion Implicit Models (DDIM)~\cite{song2022denoising} sampler with \(N = 200\) steps during inference.
Our model has ~305M parameters in the LDM and ~128M in the VAE and vocoder, comparable to baseline models.

We evaluated the separation task using the Mel mean square distance, as SI-SDR~\cite{LeRoux2018SDRH} is unsuitable for our model. The use of HiFi-GAN introduces phase reconstruction differences, preventing direct waveform comparison.
For generative tasks, we used the Frechet Audio Distance (FAD)~\cite{kilgour2019frechet} metric, a widely recognized benchmark in music evaluation. Following MSDM, in arrangement generation, we used the FAD calculation protocol from \cite{donahue2023singsong}, where generated tracks are mixed with originals of given tracks and the FAD score is calculated for the resultant mixtures.




\begin{table}[t]
    \centering
    \caption{Separation performance with MSE (on Mel-spectrogram) for all tracks (B: Bass, D: Drums, G: Guitar, P: Piano). Our model is compared with our baseline, MSDM.}
    \resizebox{0.65\columnwidth}{!}{
        \begin{tabular}{l|cccc}
            \toprule
            & \multicolumn{4}{c}{MSE (on Mel)$\downarrow$} \\
            \midrule
            Model               & B         & D         & G         & P \\
            \midrule
            VAE                  & 0.28      & 0.16      & 0.89      & 0.54 \\
            \midrule
            MSDM  \cite{mariani2024multisource}    & 6.50      & 3.20      & 12.16     & 9.11\\

            MSG-LD (w=1) & 1.86 & 1.33 & 5.79 & 3.07 \\
            MSG-LD (w=2) & \textbf{1.52} & \textbf{1.01} & \textbf{4.35} & \textbf{2.50} \\
            \bottomrule
        \end{tabular}
    }
    \label{tab:separation}
\end{table}

\section{Experiments and Results} 

\subsection{Source Separation}


For the source separation task, we experimented with a CFG weight \(w \geq 1.0\) during inference of our model. For baseline comparison, we used the publicly available pretrained model from MSDM and performed separation using the Dirac algorithm, applying the best-performing hyperparameters as reported in their paper. The resulting audio was resampled, and Mel-spectrograms were extracted for direct comparison with our metrics. As reported in Table~\ref{tab:separation}, our model significantly outperforms MSDM in the separation task. Although we acknowledge that setting \(w > 1\) may push the generated samples out of distribution, we empirically observed slightly improved performance with \(w = 2.0\), which we report in the table. We also tested \(w > 2.0\) and found that it decreased the quality of the generated samples. The VAE results in the table's top row benchmark the quality of audio for our model, showing sources processed through the VAE against the original tracks. 

\subsection{Music Generation}


For the total generation task, we generated audio tracks unconditionally with \(w = 0\), mixed them, and calculated the FAD distance between them and Slakh2100 test set mixtures. As reported in Table \ref{table:FAD_total_generation}, our model significantly outperforms MSDM, reducing FAD scores from $6.55$ to $1.36$. For additional context, we also included the best-performing MusicLDM scores from \cite{chen2023musicldm}, though it is important to note that a direct comparison is not possible due to the use of different datasets.

Additionally, we explored the setting \(0 < w < 1\), expecting softly conditioned audio generation, potentially producing audios similar to the mixture but not identical. However, the outputs were mostly noisy variations of separation, so we did not report these results and leave this direction of work for future exploration.

\begin{table}[t]
    \centering
    \caption{FAD score comparison for total music generation task. Our model in unconditional mode is compared with The MSDM baseline, along with a standard MusicLDM model.}
    \vspace{0.1cm}
    \resizebox{0.4\textwidth}{!}{
        \begin{tabular}{l|c|c|c}
        \toprule
        Model & MSDM \cite{mariani2024multisource} &  MSG-LD \(w=0\) &  MusicLDM \cite{chen2023musicldm}\\
        \midrule
        FAD $\downarrow$ & 6.55 & \textbf{1.30} & 1.68 \\
        \bottomrule
        \end{tabular}
    }
    \label{table:FAD_total_generation}
    \vspace{-10pt} 
\end{table}

Table~\ref{table:arrange_generation} shows the arrangement generation experiment results. We provided our model with a subset of tracks and tasked it, with \(w = 0\), to generate the remaining ones. We conducted 14 experiments, generating all possible stem combinations, and compared our results with MSDM. Our model outperforms MSDM in every combination except for guitar stem generation. Notably, our model demonstrates weaker performance when required to add drums to the arrangement. This is evident in the slightly worse scores in the corresponding combinations. Based on informal listening, we observed that sometimes when drums are not provided, the model struggles to maintain rhythmic coherence with the given tracks, likely due to the lack of clear rhythmic cues.

\section{Conclusion}

We proposed the MSG-LD model, a versatile framework that unifies source separation and music generation tasks within the LDM paradigm. With conditioning, MSG-LD operates as a source separation model, while without conditioning, it functions as a generative model capable of tasks such as total music generation and arrangement generation. Our experiments and evaluations demonstrate that MSG-LD successfully performs all three tasks, achieving musical coherence and significantly outperforming the baseline. We acknowledge the audio quality limitations of our model, stemming from the lower 16 kHz sampling rate and the use of the original MusicLDM model's pretrained components, such as the VAE and vocoder.
Future work should explore using higher sampling rates and potentially replacing the VAE and vocoder with more advanced variations to improve latent space representation and audio quality. 
Another direction is to explore soft conditioning as a middle ground between separation and unconditional generation, allowing more control over similarity to the original track and enabling additional use scenarios, making the model more versatile.




\bibliographystyle{IEEEtran}
\bibliography{msg-ld}

\end{document}